\documentclass{iaus}
\usepackage{graphicx}

\title[Evolution of a SN Driven ISM] %% give here short title %%
{Dynamical Evolution of a Supernova Driven Turbulent Interstellar Medium}

\author[Breitschwerdt \& Avillez]   %% give here short author list %%
{Dieter Breitschwerdt$^1$ \and Miguel A. de Avillez$^{1,2}$}
\affiliation{$^1$Institut f\"ur Astronomie, Universit\"at Wien,
T\"urkenschanzstra{\ss}e 17, A-1180 Vienna, Austria \break email: breitschwerdt@astro.univie.ac.at\\[\affilskip]
$^2$Department of Mathematics, University of \'Evora, R. Rom\~ao
Ramalho 59, 7000 \'Evora, Portugal; email:
mavillez@astro.univie.ac.at}

\pubyear{2006}
\volume{237}  %% insert here IAU Symposium No.
\pagerange{119--126}
\date{?? and in revised form ??}
\jname{Triggered Star Formation in a Turbulent ISM}
\editors{B. G. Elmegreen \& J. Palous, eds.}
\begin{document}

\maketitle

\begin{abstract}
It is shown that a number of key observations of the Galactic ISM
can be understood, if it is treated as a highly compressible and
turbulent medium, energized predominantly by supernova explosions
(and stellar winds). We have performed extensive numerical high
resolution 3D hydrodynamical and magnetohydrodynamical simulations
with adaptive mesh refinement over sufficiently long time scales to
erase memory effects of the initial setup. Our results show, in good
agrement with observations, that (i) volume filling factors of the
hot medium are modest (typically below 20\%), (ii) global pressure
is far from uniform due to supersonic (and to some extent
superalfv\'enic) turbulence, (iii) a significant fraction of the
mass ($\sim 60$\%) in the warm neutral medium is in the thermally
unstable regime ($500 < {\rm T} < 5000$ K), (iv) the average number
density of O{\sc vi} in absorption is $1.81 \times 10^{-8} \, {\rm
cm}^{-3}$, in excellent agreement with Copernicus and FUSE data, and
its distribution is rather clumpy, consistent with its measured
dispersion with distance.
\end{abstract}

\firstsection % if your document starts with a section,
\section{Introduction}

Low resolution observations of the interstellar medium (ISM) at
various wavelengths reveal a rather smooth spatial distribution of
the gas and magnetic fields, and distinctive gas phases can be
discerned. Theoretical studies during the last three decades have
culminated in a widely accepted multiphase ``standard model'' (e.g.\
McKee \& Ostriker 1977 (MO-model), McKee 1990), in which the gas is
distributed into three phases in global pressure equilibrium, a cold
and warm neutral phase (CNM and WNM, respectively), a warm ionized
(WIM) and a hot intercloud (HIM) medium. There is global mass
balance by evaporation and condensation, and energy balance between
supernova (SN) energy injection and radiative cooling. One of its
testable predictions is a large volume filling factor of the HIM
($f_h \geq 0.5$) for the Galaxy. Observations, however, also in
external galaxies ($f_h \sim 0.1$, e.g.\ Brinks \& Bajaja 1986),
point to much lower values. This discrepancy can be removed if one
allows for a fountain flow due to the break-out of superbubbles into
the galactic halo (so-called chimney model, Norman \& Ikeuchi 1989)
as well as buoyant outflow from supernova remnants (SNRs). O{\sc vi}
absorption line column densities, which were pivotal in establishing
the HIM in the first place, are thought to arise in conductive
interfaces, yielding systematically too large values by an order of
magnitude, when compared to Copernicus observations (Jenkins \&
Meloy 1974). Furthermore, Jenkins \& Tripp (2006, these proceedings)
have measured C{\sc i} absorption lines towards a sample of $\sim
100$ stars from the HST archive and find a large variation in the
CNM pressure of $500 < P/k_B < 4000 \, {\rm cm}^{-3} \, {\rm K}$, in
contrast to what is expected from a model where pressure equilibrium
is a key element. Although turbulence is recognized to play an
important role in steady-state multiphase models, it is largely
treated as an additional pressure source, ignoring its
\emph{dynamical} importance.

A fundamentally different and more physical approach to model the
structure and evolution of the ISM goes back to the ideas of von
Weizs\"acker (1951) who suggested that the ISM is essentially a
highly turbulent and compressible medium. Indeed, high resolution
observations of the ISM show structures on \emph{all scales} down to
the smallest resolvable ones, implying a dynamical coupling over a
wide range of scales, which is a main characteristic of a turbulent
flow with Reynolds numbers of the order of $10^5 - 10^7$ (cf.\
Elmegreen \& Scalo 2004). Another characteristic of widespread ISM
turbulence is its enhanced mixing of fluid elements, which,
unlike thermal conduction, is largely independent of strong
temperature gradients and magnetic fields. Recently, the dynamical
importance of turbulence in the ISM and in star formation in
molecular clouds has been recognized by several groups using
different numerical approaches (e.g.\ Korpi et al. 1999,
V\'azquez-Semadeni et al. 2000, Avillez \& Breitschwerdt 2004).

Physically the generation of 3D turbulence is intimately related to
vortex stretching and its subsequent enhancement, in contrast to 2D
where vorticity is conserved. A natural way to generate vorticity is shear
flow in which transverse momentum is exchanged between neighbouring
fluid elements. This typically occurs when a flow is decelerated at
a surface (giving rise to a boundary layer) like wind gushing down a
street along the wall of a high building, or in case of
the ISM, colliding gas flows, like e.g hot gas breaking out of an
SNR or superbubble (SB). Various sources of
turbulence for the ISM have been identified: stellar (jets, winds,
H{\sc ii} regions, SN explosions), galactic rotation, self-gravity,
fluid instabilities (e.g.\ Rayleigh-Taylor, Kelvin-Helmholtz),
thermal instability, MHD waves (e.g.\ due to cosmic ray streaming
instability), with SNe representing energetically the most
importance source (see e.g.\ MacLow \& Klessen 2004).

We will show in the following sections that the new approach of a
turbulent SN driven ISM can reproduce many key observations (Avillez
2000, Avillez \& Breitschwerdt 2004, 2005a,b, henceforth AB04,
AB05a, AB05b), such as a low volume filling factor of the HIM, large
pressure fluctuations in the ISM, observed O{\sc vi} absorption
column densities by Copernicus and FUSE, and WNM gas in thermally
unstable temperature ranges.

\section{Model setup}\label{sec:model}

We have performed both hydrodynamical (HD) and magnetohydrodynamical
(MHD, with a total field of 4.5 $\mu$G, with the mean and random
components of $B_{u}=3.1$ and $\delta B=3.2$ $\mu$G, respectively)
simulations to study by adaptive mesh refinement simulations the
global and local evolution of the SN driven ISM. We use a grid
centred at the solar circle with a square disk area of 1 kpc$^{2}$
and extending from $z=-10$ to $+10$ kpc in the directions
perpendicular to the Galactic midplane. The finest resolution is
1.25 pc (MHD) and 0.625 (HD), respectively. Gravity is provided by
the stars in the disk, radiative equilibrium cooling assuming solar
and also 2/3 solar abundances (hence, log (O/H)=-3.07 (Anders \&
Grevesse 1989) and -3.46 (Meyer 2001), respectively), uniform
heating due to starlight varying with $z$ (cf.\ Wolfire et al. 1995)
and a magnetic field (setup at time zero assuming equipartition) for
the case of MHD runs. SNe types Ia and Ib+c+II are the sources of
mass, momentum and energy. SNe Ia are randomly distributed, while
the other SNe have their high mass progenitors generated in a
self-consistent way according to the mass distribution in the
simulated disk (with roughly 60\% exploding in associations) and are
followed kinematically according to the velocity dispersion of their
progenitors. In these runs we do not consider heat conduction, as
turbulence provides the dominant mixing process. For setup and
simulation details see AB04 and AB05a.

\section{Results}\label{sec:results}

In the following we focus on those results of our simulations, which
show a clear deviation from the aforementioned standard picture of
the ISM.
%which is characterized by a distribution of the
%interstellar gas into distinct phases in pressure equilibrium, with
%the HIM in the disk covering 50\% or more by volume.
It is important to emphasize that the computational box has to be sufficiently
large in order to avoid significant mass loss, and that the
evolution time is long enough (400 Myr in our runs) in order to be insensitive
of the (necessarily artificial) initial setup and to attain global dynamical equilibrium.
In addition we have checked that the results
are resolution independent by doubling the resolution and found that
changes are less than a few percent.

\subsection{Volume filling factors}
A first striking feature of global ISM simulations in a SN driven
ISM is the \textit{continuous} distribution of the plasma over
temperature, rather than in distinct phases (for a discussion see
below). We have therefore specified temperature regimes that
correspond to the classical phases as well as to thermally unstable
regimes. The second striking feature is the low volume filling
factor of the HIM (see Table~\ref{Avervff}).
\begin{table}
\centering \caption{Average volume filling factors of the different
ISM phases for variable SN rate $\sigma$ (in units of the Galactic
rate). The average was calculated using 101 snapshots (of the 1.25
resolution runs) between 300 and 400 Myr of system evolution with a
time interval of 1 Myr. \label{Avervff} }

\begin{tabular}{ccccc}
%\hline
\hline
$\sigma$$^a$ & $\langle f_{v, cold}\rangle$$^b$ &
$\langle f_{v, cool}\rangle$$^c$ &  $\langle f_{v, warm}\rangle$$^d$ & $\langle f_{v, hot}\rangle$$^e$ \\
\hline
1 & 0.171 & 0.354 & 0.298 & 0.178 \\
2 & 0.108 & 0.342 & 0.328 & 0.223 \\
4 & 0.044 & 0.302 & 0.381 & 0.275 \\
8 & 0.005 & 0.115 & 0.526 & 0.354 \\
16 & 0.000 & 0.015 & 0.549 & 0.436 \\
\hline
\multicolumn{5}{l}{$^a$ SN rate in units of the Galactic SN rate.}\\
$^b$ $T<10^{3}$ K, &
$^c$ $10^{3} <T\leq 10^{4}$ K, & $^d$ $10^{4} <T\leq 10^{5.5}$ K, & $^e$ $T> 10^{5.5}$ K. \\
%\multicolumn{5}{l}{$^a$ SN rate in units of the Galactic SN rate.}\\
%\multicolumn{5}{l}{$^b$ $T<10^{3}$ K.}\\
%\multicolumn{5}{l}{$^c$ $10^{3} <T\leq 10^{4}$ K.}\\
%\multicolumn{5}{l}{$^d$ $10^{4} <T\leq 10^{5.5}$ K.}\\
%\multicolumn{5}{l}{$^e$ $T> 10^{5.5}$ K.}\\
\end{tabular}
\end{table}
This is a result of the unavoidable setup of the Galactic fountain,
as the overpressured flow always chooses the path of least
resistance.  Even for a star formation rate, which is 16 times the
Galactic value, the HIM covers less than 50\% of the disk volume. It
should be stressed that this result is fairly robust and does not
depend strongly on the magnetic field as our MHD runs show. Even an
initially disk parallel field cannot prevent break-out, as in 3D it
is easier to push field lines aside than working against tension
forces all the way up into the halo as it is in 2D.

\subsection{The myth of pressure equilibrium}
It has been often argued that there should exist global pressure
equilibrium between the various stable phases. This hypothesis would
be correct, if there would be sufficient time for relaxation for the
various processes responsible for mass and energy exchange like
collisional heating, radiative cooling, condensation and evaporation
etc..
\begin{figure*}[thbp]
  \centering
  \includegraphics[width=0.33\hsize,angle=-90]{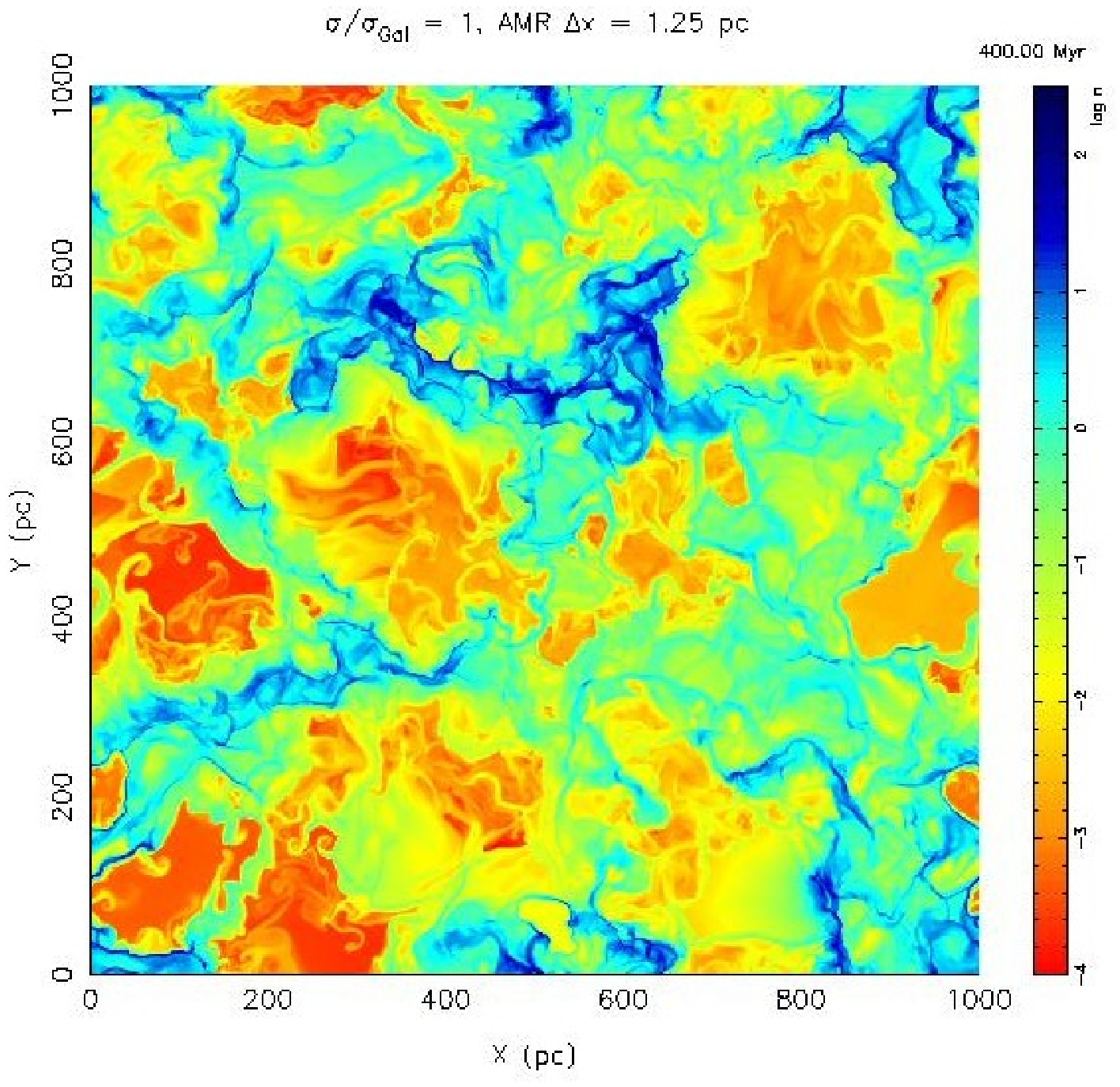}\includegraphics[width=0.33\hsize,
  angle=-90]{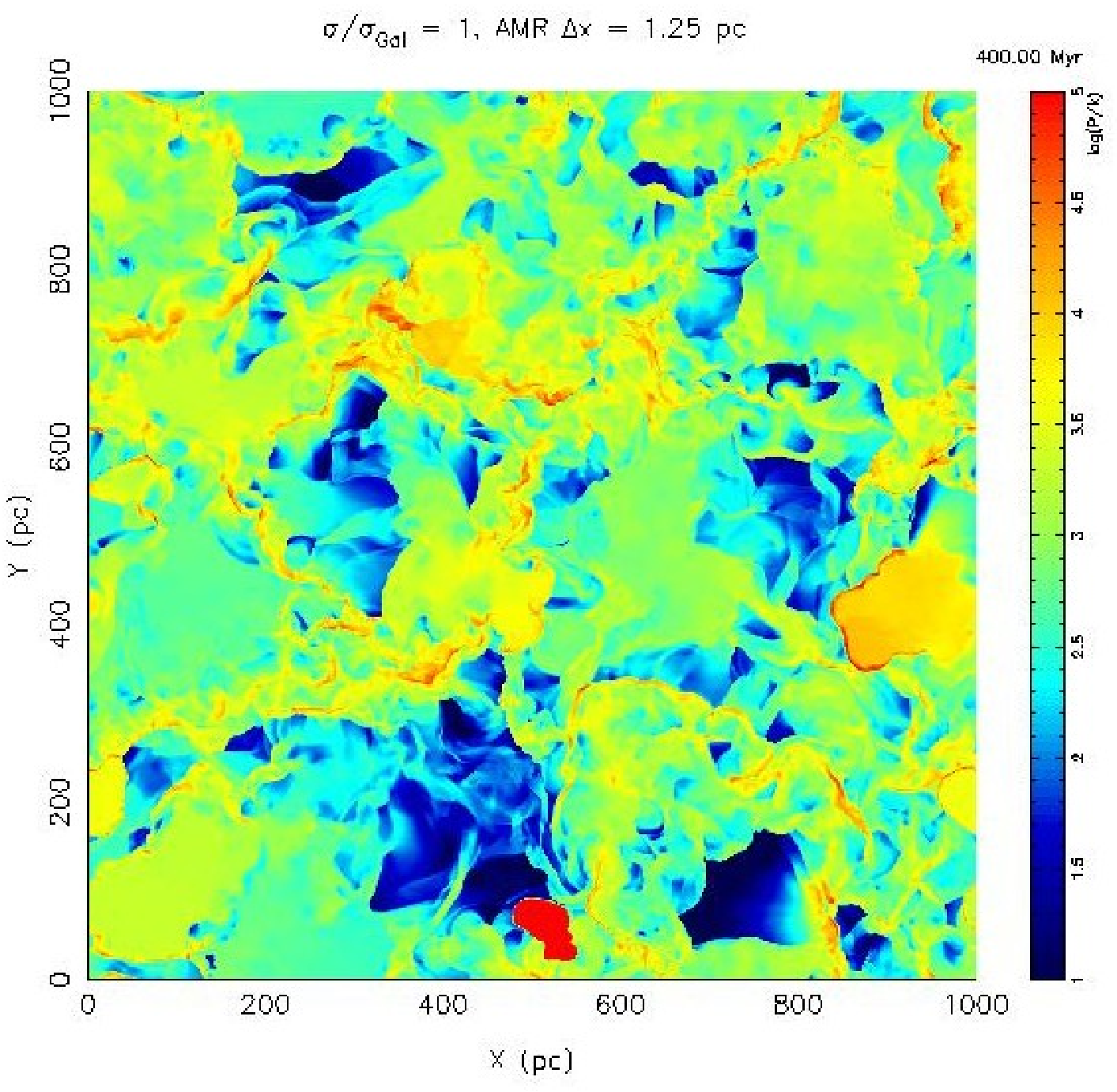}\includegraphics[width=0.33\hsize,angle=-90]{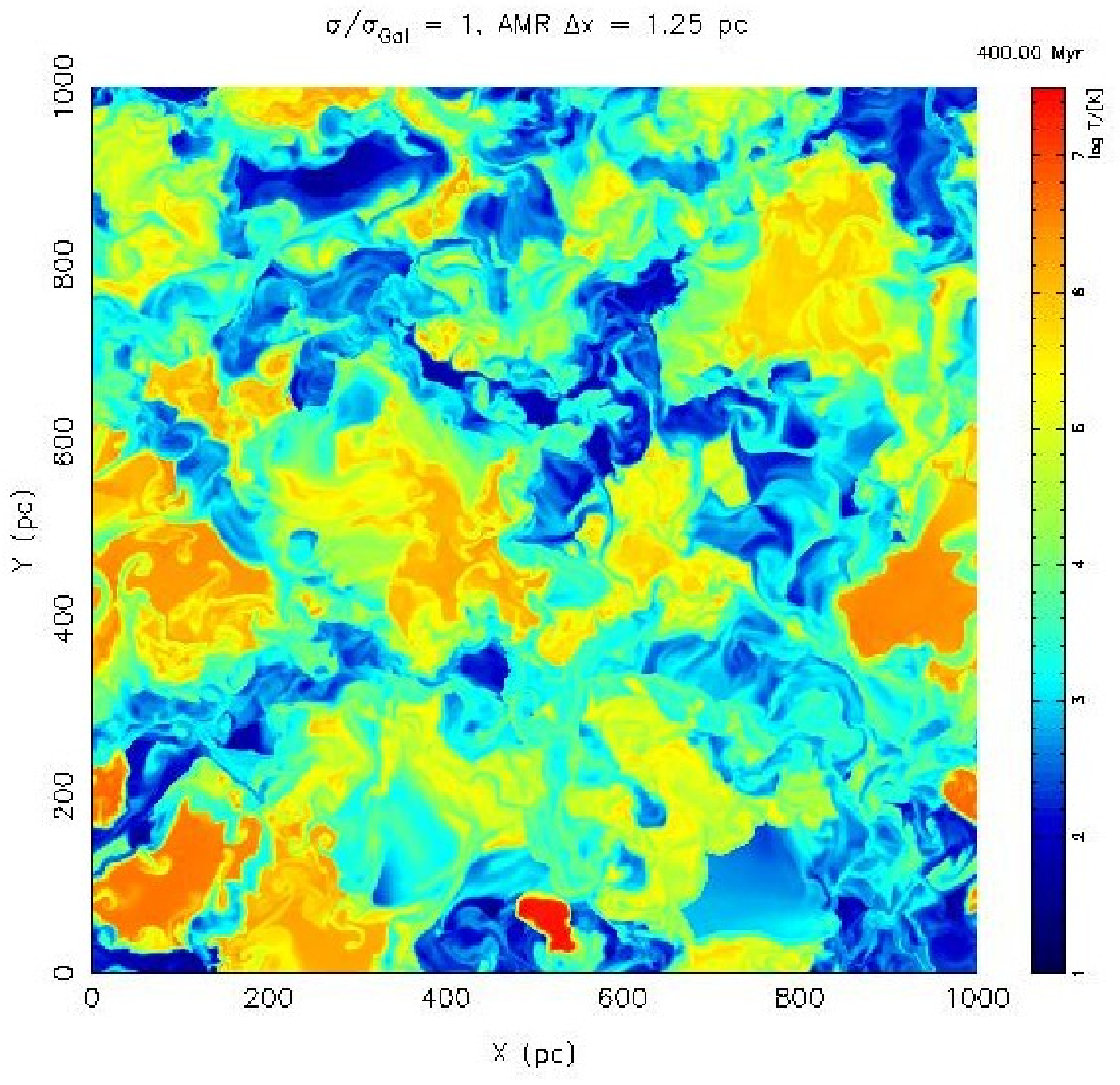}
\caption{Two dimensional cuts, through the 3D data cube, showing the density $n$ (left panel),
the pressure $P/k$ (middle panel) and the temperature $T$ (right panel) distribution in the Galactic plane
for an HD simulation with the Galactic SN rate.
\label{1.25pc-GP}}
\end{figure*}
However, due to the large Reynolds number of the flow, turbulent
mixing is the dominant exchange process, and a fortiori this occurs
supersonically in a compressible medium. Hence there is in general
not enough time to establish pressure equilibrium by pressure waves
propagating back and forth. There exists though a global
\textit{dynamical} equilibrium, depending on the boundary conditions
(e.g.\ SN rate, gravitational and external radiation field), which
results in an ``average pressure'', however with huge fluctuations
as can be seen in Fig.~\ref{1.25pc-GP}. The fact that the dynamical
evolution of the ISM is indeed governed by turbulence may be
appreciated by noting that in Fig.~\ref{1.25pc-GP} structures occur
on \textit{all scales}. This may on the other hand cast some doubt
on the results, as surely structures will occur below the resolution
limit. As our resolution checks have shown, this does not seem to be an issue here,
since the processes we describe here either dominate on larger scales or
do not exhibit any significant energy feedback from smaller to larger scales (as
might actually be the case in strong MHD turbulence).

\subsection{Does some interstellar gas reside in thermally unstable phases?}
H{\sc i} Arecibo Survey observations by Heiles \& Troland (2003)
have shown that about 48\% of the WNM can be found in the thermally
unstable regime between $500$ K and $5000$ K, and that CNM
linewidths are in agreement with supersonic turbulent motions in sheetlike
(aspect ratios of up to 280) clouds. Taken at face value this strongly supports a picture in which
clouds are immersed in a turbulent medium and are deformed by
vortex stretching as well as by shock compression.
Our simulations show that ISM turbulence can
drive and sustain turbulence inside clouds, which is alleviated to
some extent by the fractal structure of clouds, resulting e.g.\ from
colliding gas flows (cf.\ Burkert 2006).
\begin{figure*}[thbp]
\centering
\includegraphics[width=0.37\hsize,angle=0]{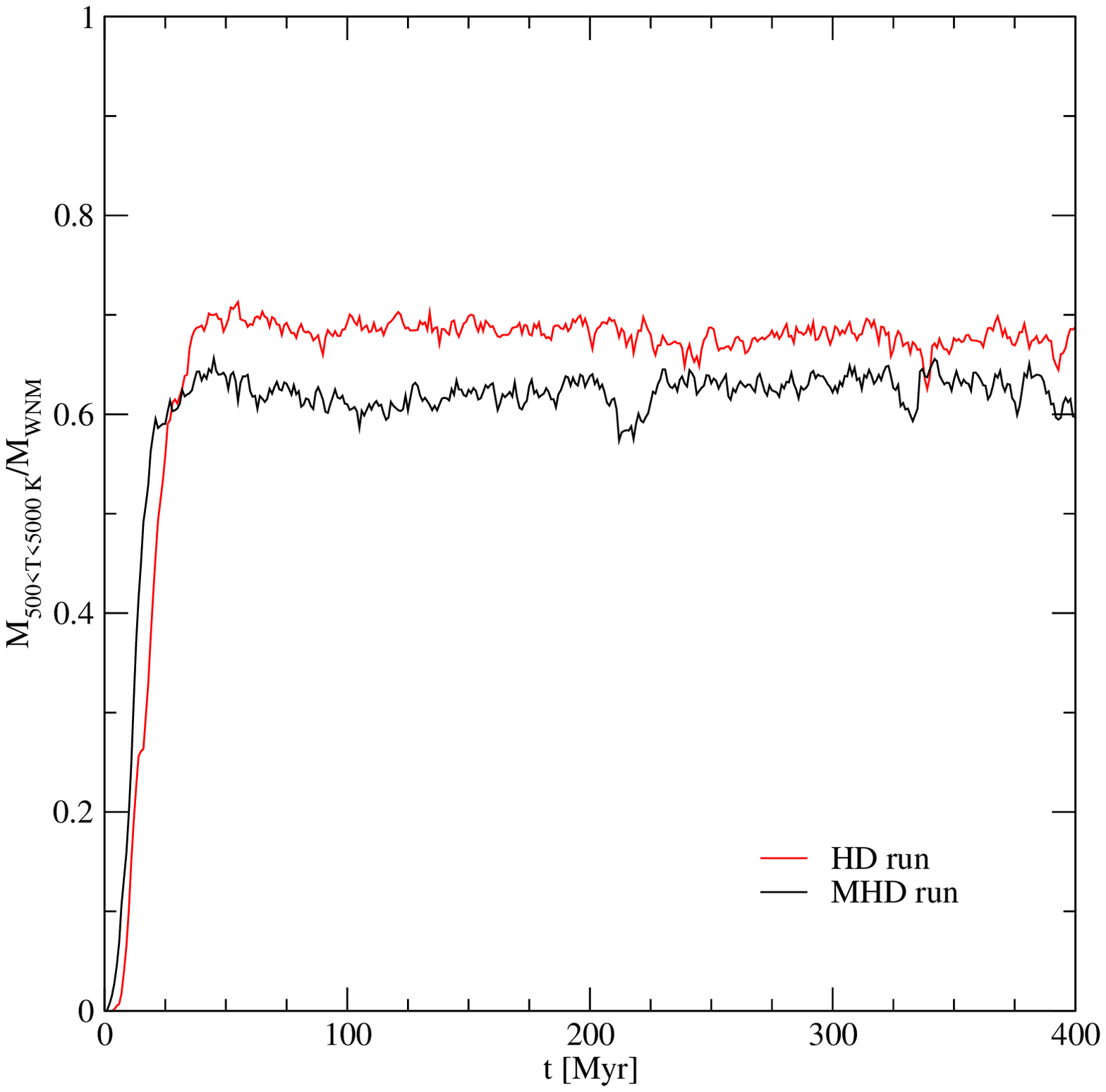}\includegraphics[width=0.40\hsize,angle=0]{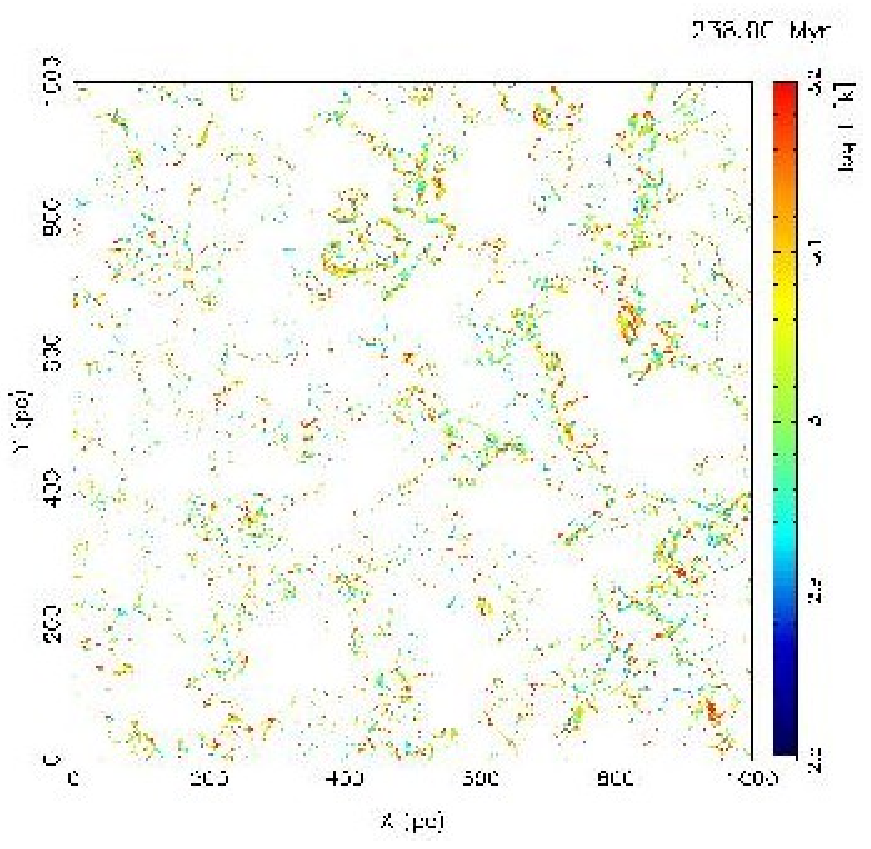}
\caption{
\textit{Left panel:} History of the fraction of mass
of the classical thermally unstable gas having $500 < T \leq 5000$
K in the disk for the HD (red) and MHD (black) runs.
\textit{Right panel:} 2D snapshot of the Galactic plane taken from run HD2a ($\Delta x=0.625$ pc) at time 238 Myr. The image shows the filamentary structure of the warm neutral gas with $2.8 \leq T \leq 3.2$ K.
\label{figmass} }
\end{figure*}
Fig.~\ref{figmass} (left) shows the fraction of the WNM derived from
our HD and MHD runs in the temperature range between $500$ and
$5000$ K, and the right panel of Fig.~\ref{figmass} demonstrates its
filamentary distribution in the narrow band between $630$ and $1590$
K. The large amount of ISM mass seen in thermally unstable regimes
is a direct consequence of SN driven turbulence. Thus, the Field
(1965) criterion is necessary, but not sufficient for distributing
the ISM gas into stable phases over different temperature ranges. It
is essential to realize that turbulence has a stabilizing effect by
inhibiting local condensation modes. The reason is that turbulence
can be regarded as a diffusion process by which energy is
efficiently transferred from large to small scales, thus preventing
thermal runaway on scales smaller than the minimal length scale over
which thermal instability can overcome turbulent diffusion (note the
small scale patchy distribution in Fig.~\ref{figmass} (right)), in
much the same way as heat conduction stabilizes the solar
chromosphere.

\subsection{Comparison of O{\sc vi} distribution to Copernicus and FUSE data}
The discovery of the widespread O{\sc vi} absorption line toward
background sources led to the discovery of the HIM (e.g.\ York 1974,
Jenkins \& Meloy 1974) and identified SNRs as a major source of hot
gas. Ever since, starting with the ``tunnel network model'' of Cox
\& Smith (1974), to reproduce the observed O{\sc vi} column
densities, N(O{\sc vi}), has been a touchstone of ISM modeling. In
collisional ionization equilibrium O{\sc vi} is the most abundant
ionization stage at $T \sim 3 \times 10^5$ K, a temperature which is
typical for transition regions between HIM and cooler gas, like e.g.\
in conductive interfaces. In the MO model these occur in
large numbers between the HIM and embedded clouds and lead
to an $n$(O{\sc vi}) number density
about an order of magnitude larger than the average value of $1.7
\times 10^{-8} \, {\rm cm}^{-3}$.
\begin{figure*}[thbp]
\centering
\includegraphics[width=0.4\hsize,angle=0]{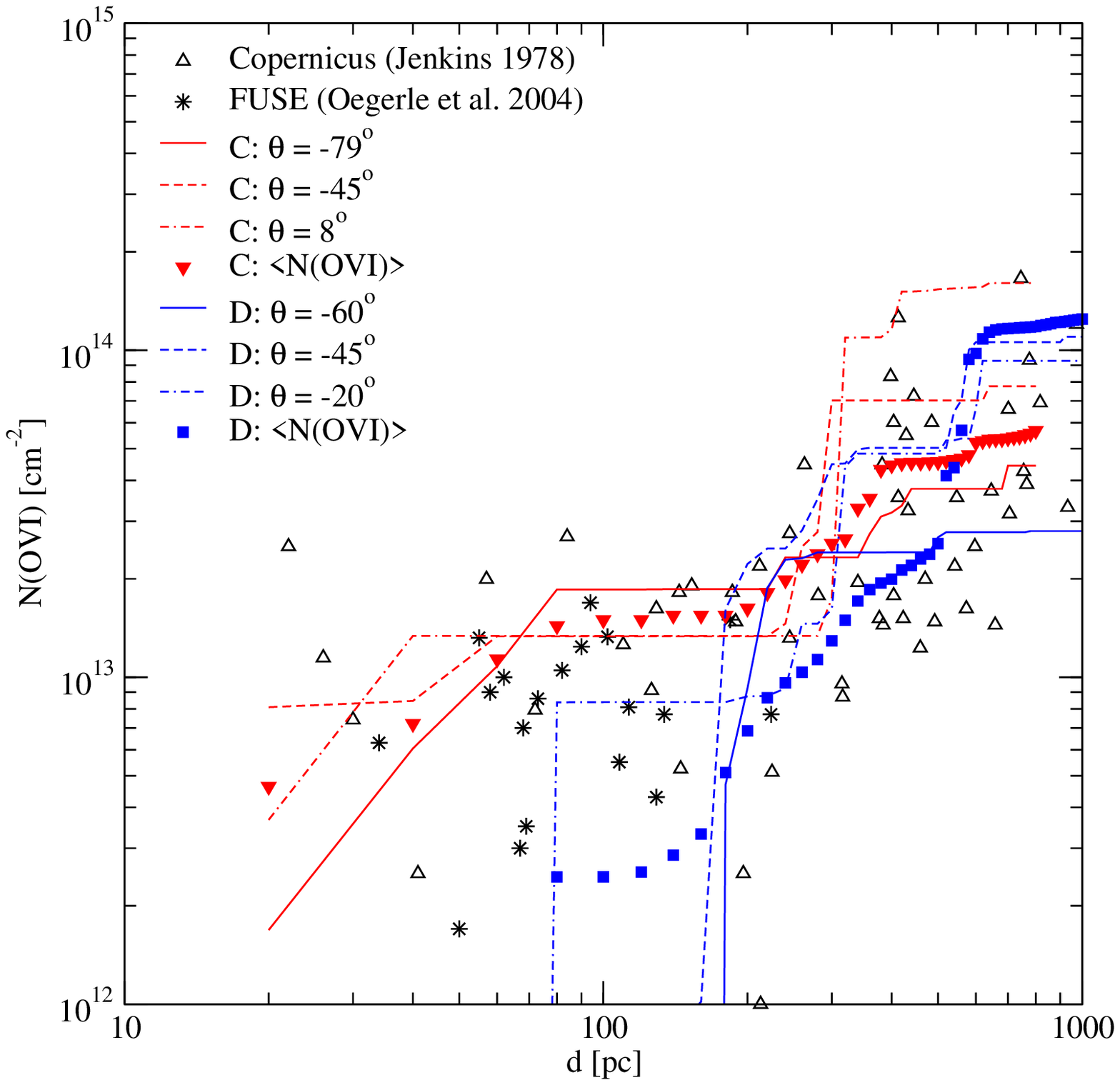}\includegraphics[width=0.6\hsize,angle=0]{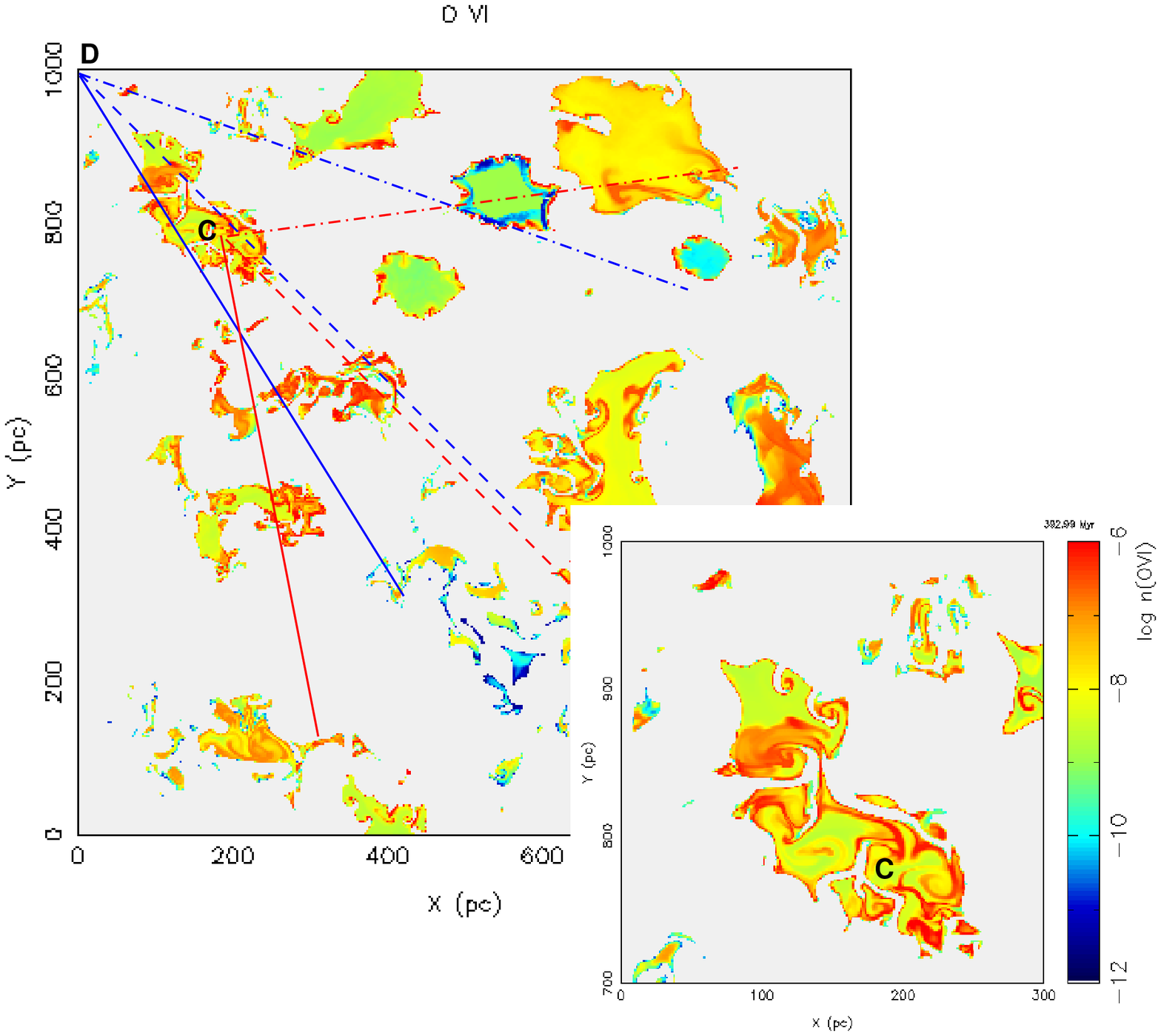}
%{mavillez_fig3b.eps}
\caption{ \emph{Left panel:} Comparison of FUSE (stars) and
\textsc{Copernicus} (open triangles) O{\sc vi} column densities with
spatially averaged (red triangles and blue squares) and single lines
of sight (red and blue lines) N(O{\sc vi}) measurements in the
simulated disk at time $t=393$ Myr. The LOS are taken at positions C
(red) and D (blue) that are located inside and outside of a bubble
cavity, respectively, as shown in the right panel. \emph{Right
panel:} O{\sc vi} density distribution (in logarithmic scale) in the
midplane at time $t=393$ Myr. The panel also includes a zoom of the
bubble located in position C. The colour scale varies between
$10^{-12}$ and $10^{-6} \, {\rm cm}^{-3}$; grey corresponds to zero O{\sc
vi}. Note the eddy-like structures of O{\sc vi} inside the bubbles.
\label{insidebubble} }
\end{figure*}
On the other hand our simulations show, that if turbulent mixing is
the dominant process in energy redistribution, the number of
interfaces is reduced, as energy transport is not primarily driven
by temperature gradients but by a turbulent cascade from larger to
smaller eddies. Fig.~\ref{insidebubble} (left) shows a comparison
between both Copernicus and FUSE data and our simulations, which is
remarkably good and yields a time averaged value of $n$(O{\sc vi}) =
$1.81 \times 10^{-8} \, {\rm cm}^{-3}$ without any tuning. Even the
measured $n$(O{\sc vi}) inside the Local Bubble is in excellent
agreement with the FUSE data. Fig.~\ref{insidebubble} (right)
stresses that the O{\sc vi} distribution occurs mainly in regions
separated by a length scale of about 100 pc. This provides an
extreme inherent clumpiness, which ensures the dispersion of the
column density to be roughly independent of distance, $d$, (for $d>
100$ pc; for details see AB05b) rather than declining with the
number $N_{cl}$ of interspersed clouds like $1/\sqrt{N_{cl}}$, fully
consistent with FUSE observations (Bowen et al. 2005).

%\begin{figure*}[thbp]
%\centering
%\includegraphics[width=0.45\hsize,angle=0]{mavillez_fig3.ps}
%\caption{History of the kinetic energy decay after switch-off of SN energy
%input at $t=200$ Myr in the different thermal regimes (solid lines)
%for run HD2b, normalized to their values at $t=200$ Myr, and the best
%fits to the data (dashed lines) using the power law $t_{Myr}^{-\eta}$
%with $\eta=0.5$, 0.55, 1.23, 1.2, and 1.18 for $T \leq 200$ K,
%$200< T \leq 10^{3.9}$ K, $10^{3.9}< T \leq 10^{4.2}$ K,
%$10^{4.2}< T \leq 10^{5.5}$ K, and $T > 10^{5.5}$ K,
%respectively, for $t\leq19$ Myr (dashed lines) and $\eta=1.12$, 1.37,
%2.57, and 2.07 for the same regimes, excluding the hot gas, at $t>19$
%Myr. The blue circles denote the rms values of the kinetic energy
%$E_{K,rms}$.
%\label{turbulent1}
%}
%\end{figure*}

\section{Conclusions}
Recent high resolution multi-wavelength observations in conjunction
with theoretical research have shown that the ISM in star forming
galaxies is a highly complex ``ecosystem''. The key to a better
understanding of its nature and evolution lies in the systematic
study of compressible HD and MHD turbulence (for more detailed
studies see Avillez \& Breitschwerdt, these proceedings). Numerical high
resolution 3D simulations offer a unique possibility to investigate
the nonlinear interaction of the physical processes at work,
together with a careful analysis of scaling laws.  Since the SN
driven ISM model can already explain many important features, as we
have shown, we feel encouraged to implement further processes, such
as e.g.\ non-equilibrium cooling, self-gravity and cosmic rays into
our bottom-up model. Since these studies require a huge amount of
massive parallel computing power, we are just at the beginning.
%to a more profound understanding of the nature of the ISM.

\begin{acknowledgments}
DB thanks Jan Palous, Bruce Elmegreen and the organizers for financial support.
\end{acknowledgments}


\begin{thebibliography}{}

\bibitem[Anders \& Grevesse]{AG89}
     {Anders, E., \& Grevesse, N.} 1989,
     \emph{Geochim. Cosmochim. Acta} 53, 197

\bibitem[Avillez (2000)]{av00}
     {Avillez, M.A.} 2000,
      \textit{MNRAS} 315, 479

\bibitem[Avillez \& Breitschwerdt (2005a)]{AB05a}
     {Avillez, M.A. \& Breitschwerdt, D.} 2005,
     \emph{A\&A} 436, 585 (AB05a)

\bibitem[Avillez \& Breitschwerdt (2005b)]{AB05b}
     {Avillez, M.A. \& Breitschwerdt, D.} 2005,
     \emph{ApJ} 634, L65 (AB05b)

\bibitem[Avillez \& Breitschwerdt (2004)]{AB04}
     {Avillez, M.A. \& Breitschwerdt, D.} 2004,
     \emph{A\&A} 425, 899 (AB04)

\bibitem[Bowen et al. (2005)]{bo05}
     {Bowen, D. V, Jenkins, E. B., Tripp, T. M. et al.} 2005, in:
      G. Sonneborn et al. (eds.),
     {Astrophysics in the Far Ultraviolet - Five years of discovery with
     FUSE}, ASPC (\textbf{astro-ph/0410008})


\bibitem[Brinks \& Bajaja (1986)]{BB86}
     {Brinks, E., \& Bajaja, E.} 1986,
     \emph{A\&A} 169, 14

\bibitem[Burkert (2006)]{Bu06}
     {Burkert, A.} 2006, in: F. Combes \& R. Robert (eds.),
     \textit{Statistical Mechanics of Non-Extensive Systems},
      Comptes Rendus Physique, vol.\ 7, p.\ 433 (\textbf{astro-ph/0605088})

\bibitem[Cox \& Smith (1974)]{cs74}
     {Cox, D.P. \& Smith, B.W.} 1974,
     \emph{ApJ} 189, L105

\bibitem[Elmegreen \& Scalo (2004]{ES04}
      {Elmegreen, B.G. \& Scalo, J.} 2004,
      \emph{ARA\&A} 42, 211

\bibitem[Field (1965)]{Fi65}
      {Field, G. B.} 1965,
       \textit{ApJ} 142, 531

\bibitem[Heiles \& Troland (2003)]{HT03}
      {Heiles, C., \& Troland, T. H.} 2003,
      \emph{ApJ} 586, 1067

\bibitem[Jenkins \& Meloy (1974)]{JM74}
     {Jenkins, E.B. \& Meloy, D.A.} 1974,
     \textit{ApJ} 193, L121

\bibitem[Korpi et al. (1999)]{Ko99}
     {Korpi, M.J., Brandenburg, A., Shukurov, A., et al.} 1999,
     \emph{ApJ} 514, L99

\bibitem[MacLow \& Klessen (2004)]{MLK04}
     {MacLow, M.-M \& Klessen, R.S.} 2004,
     \emph{Rev.\ Mod.\ Phys.} 76(1), 125

\bibitem[McKee (1990)]{MK90}
     {McKee, C.F.} 1990, in:
     \emph{The evolution of the interstellar medium}, ASPC, p.\ 3

\bibitem[McKee \& Ostriker (1977)]{MO77}
     {McKee, C.F. \& Ostriker, J.P.} 1977,
     \textit{ApJ} 218, 148

\bibitem[Meyer (2001)]{M01}
     Meyer, D.M. 2001, in: R. Ferlet et al. (eds.),
    \textit{Gaseous Matter in Galaxies and Intergalactic Space}', Editions Fronti\'eres, Paris, p.~135

\bibitem[Norman \& Ikeuchi (1989)]{NI89}
     {Norman, C. A., \& Ikeuchi, S.} 1989,
     \emph{ApJ} 345, 372

%\bibitem[Spitzer (1990)]{Sp90}
%     {Spitzer, L.jr.} 1990,
%     \emph{ARA\&A} 28, 71

\bibitem[V\'azquez-Semadeni et al. (2000)]{VS00}
     {V\'azquez-Semadeni, E., Gazol, A., Scalo, J.} 2000,
     \emph{ApJ} 540, 271

\bibitem[Wolfire et al. (1995)]{Wo95}
     {Wolfire, M. G., McKee, C. F., Hollenbach,et al.} 1995,
     \textit{ApJ} 443, 152

\bibitem[York (1974))]{Yo74}
     {York, D.G.} 1974,
     \textit{ApJ} 193, L127

\bibitem[von Weizs\"acker (1951)]{vW51}
     {von Weizs\"acker, C.F.} 1951,
     \textit{ApJ} 114, 165

\end{thebibliography}
\end{document}